\documentclass[10pt,journal,compsoc]{IEEEtran}
\usepackage{comment}
\usepackage[italian, main=english]{babel}
\usepackage{caption}
\usepackage{adjustbox}
\usepackage{algorithm}
\usepackage{graphicx}
\usepackage{listings}
\usepackage{hyperref}
\usepackage{multirow}
\usepackage{multicol}
\usepackage{setspace}
\usepackage{tabularx}
\usepackage{color}
\usepackage{float}
\usepackage{wrapfig}
\usepackage{booktabs}
\usepackage{array}
\usepackage{vcell}
\usepackage{changepage}
\usepackage{subfig}

\usepackage{tikz}
\usetikzlibrary{shapes,arrows,positioning}
\usepackage{pgfplots}
\pgfplotsset{compat=newest}

\usepackage{amsmath,amssymb,amsfonts}
\usepackage[flushmargin]{footmisc}
\usepackage{listings}
\usepackage{bm}
\usepackage[inline]{enumitem}

\usepackage{algpseudocode}
\usepackage{algorithm}
\usepackage{adjustbox}
\algnewcommand\algorithmicforeach{\textbf{for each}}
\algdef{S}[FOR]{ForEach}[1]{\algorithmicforeach\ #1\ \algorithmicdo}

\newcolumntype{P}[1]{>{\centering\arraybackslash}p{#1}}

\newboolean{showcomments}
\setboolean{showcomments}{true}         

\ifthenelse{\boolean{showcomments}}
  {\newcommand{\nb}[2]{
  \fbox{\bfseries\sffamily\scriptsize#1}
     {\sf\small$\blacktriangleright$\textit{\textcolor{red}{#2}}$\blacktriangleleft$}
   }
  }
  {\newcommand{\nb}[2]{}
   
  }

\definecolor{codegreen}{rgb}{0,0.6,0}
\definecolor{codegray}{rgb}{0.5,0.5,0.5}
\definecolor{codepurple}{rgb}{0.58,0,0.82}
\definecolor{backcolour}{rgb}{0.95,0.95,0.92}

\lstdefinestyle{mystyle}{
    backgroundcolor=\color{backcolour},   
    commentstyle=\color{codegreen},
    keywordstyle=\color{magenta},
    numberstyle=\scriptsize\color{codegray},
    stringstyle=\color{codepurple},
    basicstyle=\ttfamily\scriptsize,
    breakatwhitespace=false,         
    breaklines=true,                 
    captionpos=b,                    
    keepspaces=true,                 
    numbers=left,
    numbersep=-7pt,                  
    showspaces=false,
    escapechar=|,
    showstringspaces=false,
    showtabs=false,                  
    tabsize=2
}

\ifCLASSOPTIONcompsoc
  \usepackage[nocompress]{cite}
\else
  \usepackage{cite}
\fi
\ifCLASSINFOpdf
\else
\fi
\begin{document}

\title{Understanding Fuchsia Security}

\author{Francesco Pagano,
        Luca Verderame,
        and~Alessio~Merlo
        

}



\IEEEtitleabstractindextext{%
\begin{abstract} 

Fuchsia is a new open-source operating system created at Google that is currently under active development.
The core architectural principles guiding the design and development of the OS include high system modularity and a specific focus on security and privacy. 
This paper analyzes the architecture and the software model of Fuchsia, giving a specific focus on the core security mechanisms of this new operating system.

\end{abstract}

}

\maketitle

\IEEEdisplaynontitleabstractindextext

%
\IEEEpeerreviewmaketitle


%

\ifCLASSOPTIONcompsoc

\ifCLASSOPTIONcaptionsoff
  \newpage
\fi

\section{Introduction}
\label{sec:introduction}

Fuchsia is an open-source operating system developed by Google that prioritizes security, updatability, and performance \cite{fuchsiaOverview} with the aim to support a wide range of devices, from embedded systems to smartphones, tablets, and personal computers. 

Fuchsia tries to overcome Google's end-user-facing operating systems, e.g., Chrome OS \cite{chromeOs}, Android \cite{android}, Wear OS \cite{wearOs}, and Google Home \cite{googleHome} by providing a unique OS for a wide range of devices and application scenarios.
The new OS first appeared in August 2016 in a GitHub repository, while the first official appearance was during the Google I/O 2019 event \cite{googleIo2019}.
To demonstrate the applicability of Fuchsia in a real context, in May 2021 Google released Fuchsia OS as a part of a beta testing program on its Nest Hub devices \cite{nestHub}.

The main purpose of the Fuchsia OS is to simplify the development of applications (hereafter, apps) on different kinds of devices by supporting multiple application environments.
In Fuchsia, a \texttt{component} is the common abstraction that defines how each piece of software (regardless of source, programming language, or runtime) is described and executed on the system \cite{component}. Developers can design apps as a set of components that can run in independent runtime environments, called Runners. A \texttt{Runner} can either execute binary files, render web pages, or run compiled code inside a virtual machine.
Finally, each app is distributed through a platform-independent archive, called \texttt{Package}, that includes all the required dependencies to compile and execute the software.

Such an approach aims to overcome the jeopardization of commercial OSes, like Android, where vendor-specific distributions and heterogeneous hardware configurations boost the complexity in the development of apps. 

Also, the architectural design goals of Fuchsia involve a security-oriented design that adopts the principle of least privilege and introduces security mechanisms such as \texttt{process and component isolation} (called \emph{sandboxing}), \texttt{namespaces}, and \texttt{capabilities routing} \cite{fuchsiaSecurity}.
Thus, each piece of software running on the system, including apps and system components, executes in a separated environment having the minimum set of capabilities required to perform its task. 


This article discusses the basic of Fuschia OS and its software model, with a specif focus on the security mechanisms introduced so far to protect both the OS and the apps. 
In detail, this paper provides: 
\begin{itemize}
\item a detailed description of the Fuchsia OS architecture and its software model by analyzing both the official documentation \cite{officialDocs} and the source code repository \cite{fuchsiaRepository}, updated on August 2021 (i.e., the release $f5$).
\item an in-depth analysis of the core security mechanisms implemented in Fuchsia. 
\item a discussion on the security posture of the OS, identifying some research opportunities that could enhance the security of the Fuchsia ecosystem or contribute to the finding of security weaknesses.
\end{itemize}

\textit{\textbf{Organization}}. 
The rest of the paper is organized as follows: Section \ref{sec:fuchsia_os} presents the Fuchsia platform architecture and the software model, while Section~\ref{sec:component_description} describes the runtime behavior of apps (and components) and their interactions within the system. 
Section~\ref{sec:securty_mechanisms} provides an in-depth analysis of the security mechanisms of Fuchsia. Finally, Section~\ref{sec:discussion} provides a discussion highlighting the strengths and possible weaknesses of the OS and points out some future research directions.

\section{The Fuchsia Platform Architecture}
\label{sec:fuchsia_os}


The Fuchsia architecture consists of a stack of layers providing the OS core functionalities, as depicted in Figure \ref{fig:fuchsia_architecture}. 
The bottom layer is the Zircon Kernel. On top of it, there live four layers that manage the hardware drivers (Drivers layer), the file system (File Systems layer), the OS runtime (Fuchsia Runtime), and the software components (Component Framework). Finally, the Application Layer comprises all the user-space apps. The rest of this section briefly describes each layer.


\begin{figure}[!ht]
    \centering
    \includegraphics[width=0.48\textwidth]{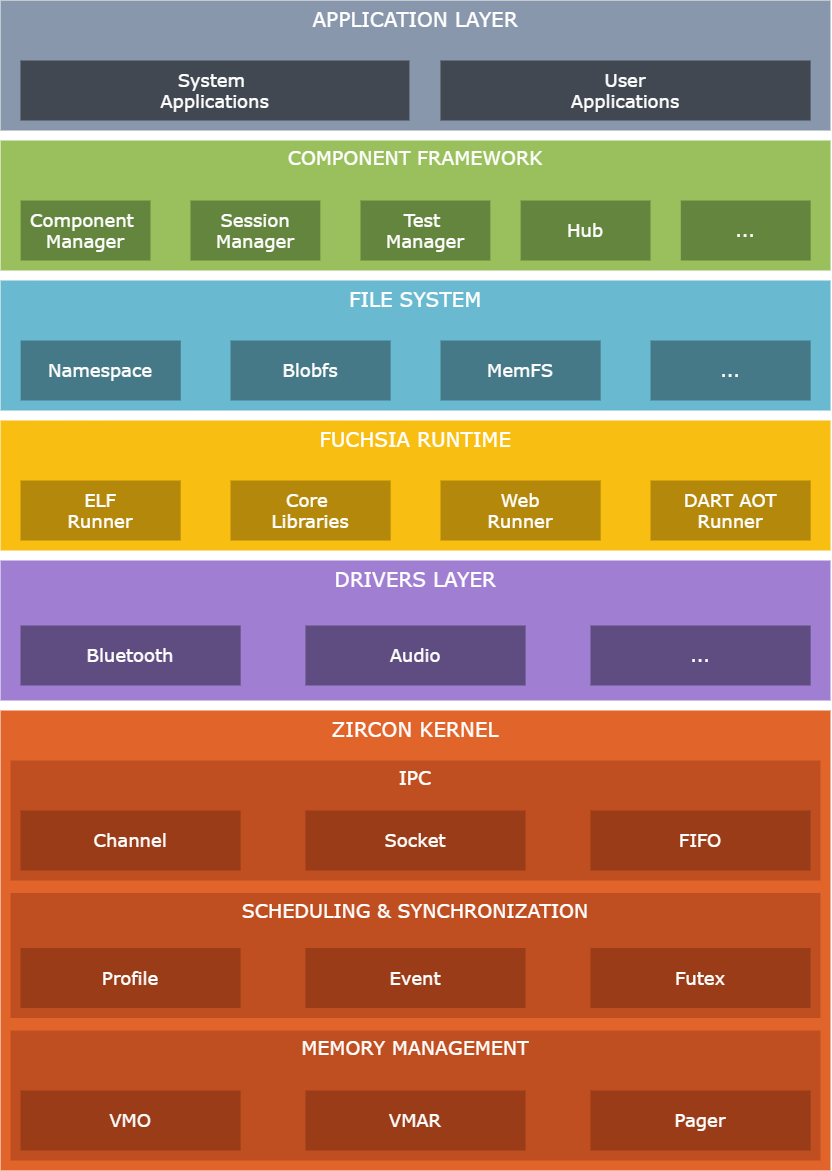}
    \caption{The architecture of Fuchsia OS.}
    \label{fig:fuchsia_architecture}
\end{figure}

\subsection{Zircon Kernel}

The core of Fuchsia OS is the Zircon kernel, a micro-kernel developed in C language and based on the Little Kernel project \cite{lk}.
Zircon is an object-based kernel, which means that all its functionalities are divided among different execution units, called \emph{objects} \cite{kernelObjects}, that are instantiated as needed. These objects are interfaces between user processes and OS resources. There is an object for every kind of task that the kernel has to manage, such as inter-process-communication (IPC), scheduling, and memory management.


\subsubsection{Inter-Process-Communication}

The Zircon kernel enables the communication between different isolated processes through the following mechanisms:

\begin{itemize}
    \item \textbf{Channels} are kernel objects that implement a bidirectional communication tunnel. A channel has two endpoints and each of them maintains a separate FIFO queue to store incoming messages. Components may use channels through the \texttt{libfdio} library that provides primitives to deliver and receive messages.

    \item \textbf{Sockets} are a bidirectional stream transports that enable processes to exchange data, using specific primitives. At the creation of the socket, the process can specify the in-bytes dimension of the socket.
    
    \item \textbf{FIFOs} are first-in-first-out interprocess mechanisms that exploit shared memory between processes to maintain a queue of messages. FIFOs are the most efficient mechanism in terms of I/O operations at the cost of a severely restricted buffer size.
\end{itemize}

\subsubsection{Scheduling and Synchronization}
The Zircon scheduler is priority-based and relies on the LK scheduler \cite{lk}, which supports up to 32 priority levels and a different priority queue for each available CPU. Also, Zircon manages real-time threads that are allowed to run without preemption and until they block, yield, or manually reschedule \cite{scheduling}.
The main kernel objects responsible for the scheduling activities are:
\begin{itemize}
    \item \textbf{Profile}. This object allows for the definition of a set of high-level scheduling priorities that the Zircon kernel could apply to processes and threads. A single profile object could be linked with one or more processes or threads. If this is the case, the scheduler will put them at the same level.
    \item \textbf{Event}.  Event objects are generated to signal specific events for synchronization among processes.
    \item \textbf{Futex}. Futex objects are low-level synchronization primitives that offer a building block for higher-level APIs such as \texttt{pthread\_mutex\_t} and \texttt{pthread\_cond\_t}. Futex objects do not have any \texttt{right} \cite{handlesRights} associated with them.
\end{itemize}


\subsubsection{Memory Management}


The Zircon kernel empowers virtual memory \cite{virtualMemory} and paging to implement memory management. The memory management task is maintained thanks to specific kernel objects that handle the virtual address space of a process and its demand for new memory pages, i.e.: 



\begin{itemize}
    \item \textbf{Virtual Memory Object (VMO)}. The VMO is an object that allows the process to interact with the underlying memory pages, and it can also provide on-demand pages when needed. It is usually used to store dynamic libraries, provide their procedures to the process, and share memory pages between processes.
    \item \textbf{Virtual Memory Address Region}. This type of object manages the virtual address space of a process and abstracts the set of its addresses. A program starts with a single VMAR that manages the entire process address space. The single VMAR can be further divided by creating new VMARs that manage different parts of the process address space. Each VMAR has a set of rights on the underlying memory. The starting VMAR holds all the rights (read, write, execution). All VMARs created by a process are related to each other by a tree data structure.
    \item \textbf{Pager}. Pager objects are responsible for supporting data provider entities, such as file systems, to cache data they have to pass to other processes. Their main task is to create VMO objects and provide them with the pages to store data.
\end{itemize}


\subsection{Drivers Layer}

Fuchsia manages the interaction with the underlying hardware using user-space drivers. Such a choice i) allows relieving the Kernel from the burden of direct management, ii) increases the portability of the OS, and iii) reduces the number of context switches between user and kernel mode.


To support the development of drivers, Fuchsia OS provides the Fuchsia Driver Framework. The Fuchsia Driver Framework groups a set of tools that enable a developer to create, test and deploy drivers for Fuchsia devices. One of the most important tools inside this framework is the driver manager \cite{driverFramework}. This process is one of the first started processes at the boot of the system. It is responsible for searching for the proper drivers for each device, running them, and managing their lifecycle. It allows processes to interact with drivers through a shared file system called device file system (or \texttt{devfs}), which is mounted by each component at \texttt{/dev} path. 
Although Fuchsia OS is a recent operating system, several vendors have started to implement their drivers; the implemented drivers are listed on the official page \cite{implementedDrivers}.


\subsection{Component Framework}
The Component Framework (CF) \cite{component} consists of the core concepts, tools, APIs, runtime, and libraries necessary to describe and run components and to coordinate communication and access to resources between components. 
The CF contains the following core services:

\begin{itemize}
    \item \textbf{Component Manager}. This process is at the core service of Fuchsia, coordinates the entire lifecycle of all components, intermediates their interactions, and manages the communication with the system resources and the underlying kernel objects. 
    It is one of the first processes created when the system boots and it is one of the last processes destroyed when the system shuts down.

    \item \textbf{Session Manager}. The Session Manager offers component capabilities (e.g., storage, network access, and hardware protocols) to the so-called Session components. A session is a component that encapsulates a product’s user experience. It is the first product-specific component started at boot. For example, the session for a graphical product instantiates Scenic (graphics) as a child component to render its UI.
    The Session Manager handles the management of user "apps" and their composition into a user experience and handling of user input data flow.
    
    \item \textbf{Test Manager}. This service is responsible for running tests on a Fuchsia device. The Test Manager exposes the RunBuilder protocol, which allows starting test suites, launched as a child of test manager. The Test Manager supports the definition of test suites in a plethora of programming languages, including C/C++, Rust, Dart, and Go. 
    
    
    \item \textbf{Hub}. The Hub provides access to detailed structural information about component instances at runtime, such as their names, job and process ids, and exposed capabilities. For example, the Component Manager relies on the Hub to obtain information regarding the relationships among components, in order to grant the access to specific resources or capabilities.
    
\end{itemize}

\subsection{Fuchsia Runtime}
Fuchsia provides different kinds of runtimes that can handle components. These runner processes interact with the Component Manager to load the proper configurations for a component to be run \cite{runners}. 
A component must specify the type of runner and the required configurations inside its manifest file. The Component Manager is responsible for submitting the starting of new components to the proper runner.
Fuchsia supports the use of generic and user-defined runtimes and it currently offers the following default runners:
\begin{itemize}
    \item \textbf{ELF Runner}: this runner is responsible for executing elf files, for example, the ones generated after the compilation phase of a C or Rust program. 
    \item \textbf{DART AOT Runner}: this runner executes a fully-fledged Dart Virtual Machine, to execute Dart executable files. 
    \item \textbf{Web Runner}: the Web runner renders web pages to the screen of the device through the Chromium web engine.
\end{itemize}

To execute the component, the runner may choose a strategy such as starting a new process for the component or isolate the component within a virtual machine.

Fuchsia also includes a set of core runtime libraries that provide essential functionalities for the new processes. The core libraries include \texttt{Libc} and \texttt{vDSO} to support the execution of components and system call invocations, and \texttt{libfdio} and \texttt{VFS} to support IPC and file systems operations, respectively. 

\subsection{File Systems Layer}
The Fuchsia file system is not linked nor loaded by the kernel, unlike most operating systems, but it entirely lives in user-mode space. In detail, file systems are simply user-space processes that implement servers that can appear as file systems and responds to RPC calls \cite{filesystem}.

As a consequence, Fuchsia may support multiple file systems without the burden to recompile the kernel each time. The kernel has no knowledge about files, directories, or filesystems, thus, relieving components to invoke system calls for each file operation.
Instead, the operations on a local file system are handled through the VFS library that provides the RPC primitives needed to interact with the different file system servers.
The current version of Fuchsia supports a set of predefined file systems that include MemFS, MinFS, Blobfs, ThinFs, and FVM.





The file system architecture allows for the definition of collections of elements, called \texttt{namespaces} \cite{namespaces}. In detail, a namespace is a per-component composite hierarchy of files, directories, sockets, services, devices, and other named objects provided to a component by its environment. 

\subsection{Application Layer}


This layer comprises all the user-space software, from system services to end-user apps. 
The software model of Fuchsia allows developers to design and develop their apps as a set of components which are considered the fundamental unit of an executable software. 
Components are composable, meaning that components can be selected and assembled in various combinations to create new components. A component and its children are referred to as a \texttt{realm} \cite{realm}.
The current version of Fuchsia supports the use of several programming languages for the development of apps that include C/C++, Rust, Dart, and Go, and offers a Software Development Kit (SDK) \cite{sdk} containing all the required libraries and tools to create and compile an app.

At its core, a component consists of the following parts:

\begin{itemize}
    \item \textbf{Component URL.} The primary use of component URLs is to uniquely identify a component in the definition of a component instance. A component URL can, in principle, have any scheme. User-space components distributed in a Fuchsia package are usually identified using the \texttt{fuchsia-pkg} schemes that specifies the repository of the package, the package name, the build variant, and the name of the manifest file. For instance, \texttt{fuchsia-pkg://fuchsia.com/stash\#meta/sta\-sh\_secure.cm} identifies the stash package located in the official fuchsia repository.
    
    \item \textbf{Component Manifest.} The manifest describes the structure of the component, such as the type of runtime to execute the component, the list of dependencies and the rules to manage the resources owned and used by component. The manifest are then compiled in a binary form that uses the \texttt{.cm} extension.
    Listing \ref{lst:example_of_component_manifest} shows the source of the component manifest file of a hello\_world app\footnote{\url{https://cs.opensource.google/fuchsia/fuchsia/+/main:examples/components/basic/meta/hello-world.cml}}. In detail the manifest specifies the type of runner (row \ref{line:runner}), the executable file (\ref{line:binary}) and the external library to include (row \ref{line:include}).
    \begin{lstlisting}[caption={Component manifest of a hello-world app.}, label={lst:example_of_component_manifest}]
    {
        include: [ "syslog/client.shard.cml" ], |\label{line:include}|
        program: {
            runner: "elf", |\label{line:runner}|
            binary: "bin/hello_world", |\label{line:binary}|
        },
    } \end{lstlisting}
    
\end{itemize}


In Fuchsia, apps and system components are distributed through packages which is a distribution unit for programs, components and services \cite{packages}. 
Packages contain the set of file, separated in a specific directory structure, that enables Fuchsia OS to provide specific programs or components.

\subsection{Application Layer}
This paragraph groups some entities that are the basis for the developing of applications on Fuchsia OS:
\begin{itemize}
    \item \textbf{Package}: packages are archives used to 
    \item \textbf{Flutter}:
    \item \textbf{Scenic}:
\end{itemize}
\end{comment}

\section{Component Runtime Execution}
\label{sec:component_description}

Fuchsia supports the execution of software components using various runtimes, such as running them using native processes or by relying on virtual machines. Such a choice enables the execution of heterogeneous executables ranging from elf files to web pages and even Android apps \cite{fuchsiaVirtualization}.
Thanks to the component abstraction, Fuchsia is capable of transparently executing the software using the CF and its runtime. In the following, we discuss the lifecycle of a component, focusing on its interaction with other processes and the underlying OS. 

\subsection{Component Lifecycle}

The Component Manager controls the creation of components objects and their transitions through the different states of their lifecycle.
Figure \ref{fig:component_lifecycle} shows the lifecycle of a component, which is composed of five states, as described below: 

\begin{figure}[!ht]
    \centering
    \includegraphics[width=0.48\textwidth]{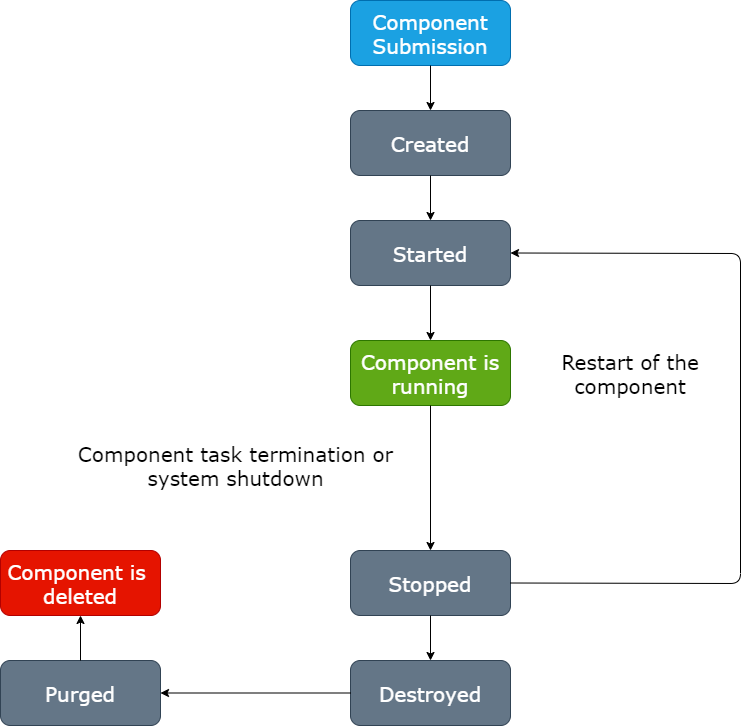}
    \caption{A simplified illustration of the Component lifecycle.}
    \label{fig:component_lifecycle}
\end{figure}

\begin{enumerate}
    \item \textbf{Created.} Upon receiving a request to start a new component, the Component Manager locates its corresponding package archive, extracts the content, and creates a \texttt{ComponentStartInfo} object containing all the necessary data to execute the component.  Such information includes the URL of the component, the type of request runner, the executable arguments, and references to kernel objects needed to start the component instance. 

    \item \textbf{Started}. After the creation of a component, the Component Manager triggers the required Runner to execute the component. The Component Manager controls the execution of a component instance by interacting with the Runner through the \texttt{ComponentController} protocol \cite{runners}.

    \item \textbf{Stopped}. Stopping a component instance terminates the component's program. Still, the Content Manager preserves the component's state by saving all the data in the persistent storage so that it can continue where it left off when subsequently restarted.
The stop event occurs if the component terminated its task or if the system is shutting down. 

    \item \textbf{Destroyed}. Once destroyed, a component instance ceases to exist and cannot be restarted. Still, it continues to exist in persistent storage.

    \item \textbf{Purged}. The Component Manager completely removes a component from the system, including all the information saved in the persistent storage.
\end{enumerate}

\subsection{Component Topology}

In Fuchsia components can assemble together to make more complex components.
At runtime, components instances, i.e., distinct embodiment of a component running, establish a set of relationships described as a \texttt{component instance tree}. Each instance contained in the component instance tree can be referred with an unique identifier called \texttt{moniker}.

A component instance (called parent or root) can build a parent-child relationship by instantiating other components, which are known as its children. Each child component instance depends entirely on the parent, and so, it lives until the parent is alive. Children can be created either statically by declaring its existence in the component manifest or dynamically, using the Realm Framework Protocol.  

Typically, an app implements its functionaries through a composition of multiple components and, thus, it has its own component instance tree. Figure \ref{fig:component_instance_tree} shows an example of a component instance tree of a sample app composed of a parent component (A) and six child components (B-G). 

\begin{figure}[!ht]
    \centering
    \includegraphics[width=0.45\textwidth]{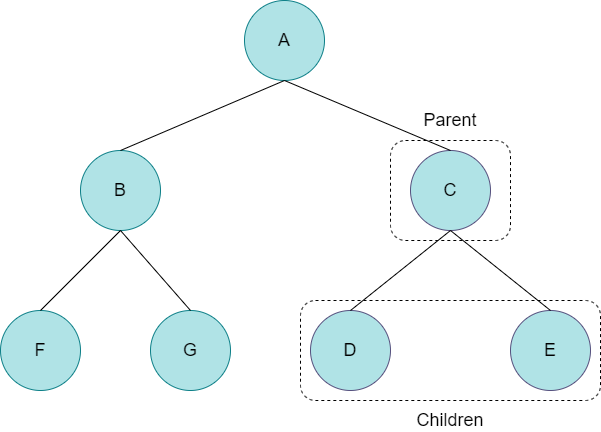}
    \caption{Example of a Component Instance Tree of a Sample App.}
    \label{fig:component_instance_tree}
\end{figure}



\subsection{Inter-Component Communication (ICC)}
\label{subsubsection:fidl}

The Fuchsia Interface Definition Language (FIDL) \cite{fidlOverview} is a language to describe how processes communicate with each other, using high-level interfaces based on inter-process-communication (IPC) mechanisms defined by the Zircon kernel.

The adoption of FIDL allows abstracting IPC between processes, and simplifies the implementation of components as it provides a standard and language-independent way to describe the interactions. 
The FIDL language allows exposing a set of functionalities (namely, a protocol) of a component (called a Service) to other processes. At runtime, the Service publishes the protocol by starting a server that responds to IPC calls made by other processes through a client interface.

The developer of a Service can create a FIDL definition file (in \texttt{.fidl} format), which describes the protocol. Then, the FIDL toolchain generates the client and server code in any of the supported target languages, as depicted in Figure \ref{fig:fidl_arch}. The generated code needs to be included in the Service and in the external components that would use the protocol.

\begin{figure}[!ht]
    \centering
    \includegraphics[width=0.45\textwidth]{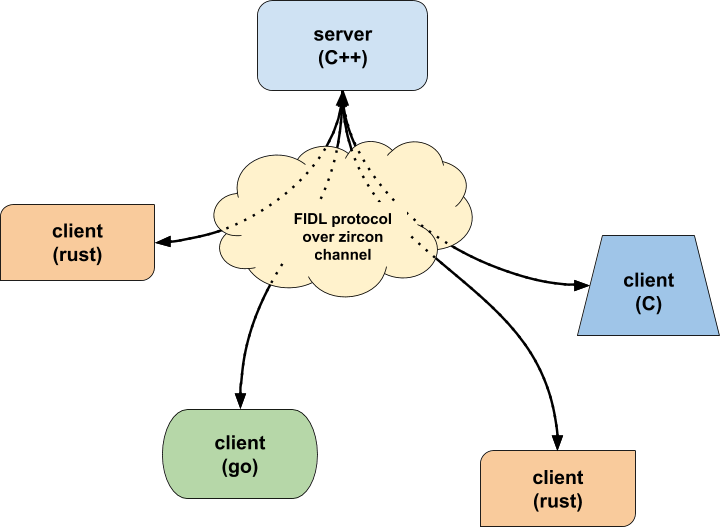}
    \caption{FIDL architecture, taken from \cite{fidlOverview}}
    \label{fig:fidl_arch}
\end{figure}

Figure \ref{lst:example_of_fidl_file} shows a snippet of a sample FIDL file. 
The example describes a protocol called Echo and a service (EchoService) that exposes an instance of Echo called \texttt{regular\_echo}. The protocol is marked as \texttt{discoverable} (row \ref{row:disco}), indicating it is available for external clients.
The Echo protocol declares a method \texttt{EchoString} that can receive a string as input (row \ref{row:input}) and respond with the same string (row \ref{row:output}).

\begin{lstlisting}[caption={Example of a FIDL file.}, label={lst:example_of_fidl_file}]
   library fuchsia.examples;

   const MAX_STRING_LENGTH uint64 = 32;

   // [START echo]
   @discoverable |\label{row:disco}|
   protocol Echo {
       EchoString(struct {
           value string:MAX_STRING_LENGTH; |\label{row:input}|
       }) -> (struct {
           response string:MAX_STRING_LENGTH; |\label{row:output}|
       });
   };
   // [END echo]

   service EchoService {
       regular_echo client_end:Echo;
   };
\end{lstlisting}

\subsection{Interaction with the OS}

User processes can interact with the underlying kernel using system calls. 
The Zircon vDSO (virtual Dynamic Shared Object \cite{vdso}) is the sole means of access to system calls in Zircon.  
The vDSO is a shared library in the ELF format provided directly by the kernel and contains all the protocols (in FIDL language) and the user-space implementation of each syscall. 
In order to use a syscall, developers need to include in their apps the \texttt{<zircon/syscalls.h>} header from the Fuchsia SDK. 

Each time a process needs to interact with a kernel object, e.g., for a system call invocation, it needs a reference to a Handle \cite{handles}.
A Handle is a kernel construct that allows user-mode programs to reference a kernel object.
Each time the Zircon kernel creates a new kernel object (as a result of a system call invocation), it returns a handle of the new object to the calling process. It is often the case that multiple processes concurrently access the same object via different handles. However, a single handle can only be either bound to a single process or to the kernel. If another process tries to use the same reference of the handle, it would access a handle that points to another object or fail because the handle would not exist.

\section{Security Mechanisms}
\label{sec:securty_mechanisms}


This section describes the main security mechanisms used by Fuchsia OS to guarantee the secure execution of components and apps, as well as the correct interaction with the kernel primitives.

\subsection{Zircon Kernel Security}

The Zircon kernel is designed to fully isolates processes by default and regulates access to its functionalities by enforcing restrictions on Objects and Handles. Also, it supports the generation of secure pseudo-random numbers and it is declared as strongly hardened against buffer overflow attacks. 

\subsubsection{Kernel Objects \& Handles Rights}

The usage of kernel objects and Handles enables Zircon to apply fine-grained control over the processes and their ability to interact with the kernel functionalities.

First, the concept of kernel object grants a separation of concerns that enables isolated and containerized kernel tasks. Also, the objects are created and disposed upon need, thus limiting the probability of interacting with unused or de-referenced objects. 

Then, user-space processes are able to interact with kernel objects only through Handles. Each Handle is bound only to a specific process, and the kernel generates it in response to a system call. Thus, a process is not expected to interact with a kernel object unless it has a valid Handle.
This approach aims to reduce the attack surface only to those objects owned by the process. 
For example, suppose an attacker gained access to a VMAR object through a Handle owned by a given process. 
In that case, she can only interact with the virtual address space of that process, but she is not expected to be able to propagate the attack on the virtual address space of other processes or the kernel. 

Finally, Handles can inherit a number of rights \cite{handlesRights} that describe the set of operations allowed on the kernel object (i.e., the system call that the process can invoke with the Handle). This mechanism allows the kernel to differentiate the access to the same object (e.g., a Channel object) on a per-process basis.
Also, even though a process can duplicate and maintain different Handles referring to the same object, each copy could have at most the same rights as the original one, granting the least privilege.

\subsubsection{Kernel Code Hardening}
The Zircon kernel is compiled using the Clang/LLVM compiler \cite{clang} with the SafeStack \cite{safeStack} and ShadowCallStack \cite{shadowCallStack} security options enabled.

SafeStack and ShadowCallStack provide the running program with two stacks: a safe stack and an unsafe one. The unsafe stack is similar to the standard stack, so it can store references to objects on the heap memory, while the safe stack is reserved for the function call stack.

Such mechanisms aim to protect the Zircon kernel against stack-smashing attacks \cite{bufferOverflow}.  In such a case, a buffer overflow attack targeting a runtime variable in the program cannot overwrite the return addresses of the function call stack since it is located in a separated memory region.
Also, Fuchsia extends such protection to all user programs developed in C by providing Clang as the default compiler.

\subsubsection{Secure Pseudo-random Number Generation}

Fuchsia OS supports the secure generation of random numbers directly by the Zircon kernel exploiting the \texttt{zx\_cprng\_draw} system call \cite{draw}. 
The generation of a random number uses the Chacha20 \cite{chacha20} cryptographic algorithm.
This algorithm requires a key and an integer nonce. The nonce passed to this algorithm is incremented after each invocation of the draw system call. 
The seed, instead, is generated by an entropy function. Unlike other operating systems, in Fuchsia OS the kernel generates entropy without relying on the device driver. Such a choice depends on the fact that drivers run in user mode and, thus, they cannot be trusted in principle. The kernel ensures a frequent update of the seed value by generating a new number every 30 minutes.

\subsection{Sandboxing}

Sandboxing is a security mechanism to isolate programs from each other at runtime. Fuchsia enforces sandboxing by executing all programs inside their isolated environment with the minimal set of rights needed to execute. 
In Fuchsia, each newly created process is isolated and does not possess any privilege, i.e., it cannot access kernel objects, allocate memory, or execute code \cite{generalSandboxing}. 
After the creation, the OS assigns the process with a minimum set of handles and resources to be allocated for the execution of a program.

Depending on the runtime environment, a program can run using different strategies, like being executed in a separated process or launched inside a separate Dart Virtual Machine hosted by a process. The component does not always correspond to a Zircon process, as documented in \cite{runners}.
Thus, Fuchsia enforces the sandboxing also at the granularity of single components.
 
\subsubsection{Namespace}
The main principle that enables component isolation is the namespace. 
Unlike other operating systems, Fuchsia does not have a "root" filesystem.
Instead, the Component Manager provides each component with one or more namespaces tailored to their specific necessities. 
The component has complete control over the assigned namespaces and runs as its own private "root" user.

The absence of users and a global root namespace makes impossible privilege escalation attacks \cite{privilegeEscalation}. 
Indeed, if an attacker takes control of a process, she would only have access to the component's resources, as Object paths (e.g., Handles and files) may not be meaningful outside the namespace boundaries.
To further enforce segregation among namespaces, Fuchsia has disabled the Dot Dot command \cite{dotdot} to prevent path traversal attacks \cite{pathTraversal}.
Components can access namespaces belonging to different components - with the granularity of a directory and its sub-directories - only through IPC mechanisms that are protected with access control mechanisms (see Sections \ref{sec:cap} and \ref{sec:cap-rout}).

\subsection{Capabilities}
\label{sec:cap}

Fuchsia is a \textsl{capability-based} operating system. A \texttt{capability} is a token defining how components can interact with system resources and other components. Only the possession of the appropriate capability enables a component to access the corresponding resource. 

Fuchsia capabilities are classified into seven types \cite{capList}, namely $Directory$, $Event$, $Protocol$, $Resolver$, $Runner$, $Service$, and $Storage$ capabilities.
Each type refers to a specific set of resources. For instance, the Directory and Storage capabilities manage access to specific directories and namespaces inside the local file system of components.

Capabilities can be defined, routed, and used in a component manifest to control which parts of Fuchsia have the ability to connect to and access which resources.

For example, if a component \texttt{A} wants to define a capability on its \texttt{data} directory, it can include in the component manifest the code of Listing \ref{lst:capability_definition}. The directory will be mounted in \texttt{/published-data} with read, write, and execute permissions.

\begin{lstlisting}[caption={Definition of a directory capability.}, label={lst:capability_definition}]
   {
       capabilities: [
            {
                directory: "data",
                rights: ["r*"],
                path: "/published-data",
            },
        ]
    }
\end{lstlisting}

\subsubsection{Capability Routing}
\label{sec:cap-rout}

The capability routing graph \cite{routing} describes how components gain access to capabilities exposed and offered by other components in the component instance tree, building a \emph{provider-consumer} relationship.

Components define the use and the sharing of capabilities that they provide in their manifest file. Such capability routes are determined by the \texttt{use}, \texttt{offer}, and \texttt{expose} declarations \cite{routingTerminology}.

The \texttt{use} declaration enables a component to require a capability in its namespace at runtime. 
Listing \ref{lst:example_of_use_declaration} shows how a component \texttt{A} can ask for a Directory capability regarding the \texttt{data} folder.

\begin{lstlisting}[caption={Example of use declaration.}, label={lst:example_of_use_declaration}]
   {
     use: [
         {
             directory: "data",
             rights: ["rw*"],
             path: "/data",
         },
     ],
  }
\end{lstlisting}

The \texttt{offer} declaration, instead, states how a component can route a particular capability to its child component instances. 
For instance, Listing \ref{lst:example_of_offer_declaration} states how the component \texttt{A} (i.e., the one that declares the capability on the \texttt{data} folder) offers the corresponding capability to its child \texttt{B}.
\begin{lstlisting}[caption={Example of offer declaration.}, label={lst:example_of_offer_declaration}]
   {
        offer: [
            {
                directory: "data",
                from: "self",
                to: [ "#B" ],
            },
        ],
    }
\end{lstlisting}

Finally, the \texttt{expose} declaration states that the component can route a given capability to its parent in the component instance tree. In such a case, the component \texttt{A} of the previous examples expose the declared directory capability to its parent.
\begin{lstlisting}[caption={Example of expose declaration.}, label={lst:example_of_expose_declaration}]
   {
        expose: [
            {
                directory: "data",
                from: "#A",
            },
        ]
    }
\end{lstlisting}

It is worth noticing that capabilities routing can happen only among visible components, i.e., that belong to the same component instance tree.

The Component Manager is responsible for managing the capability routing process and manage the component instance tree for all apps. 
When a component requests access to a resource provided by another component, it forwards the request to the Component Manager to establish whether it can hold the appropriate capability. To do so, the manager navigates the component instance tree to check whether: i) the target component is reachable (i.e., it is included in the component instance tree), and ii) it exists a path between the caller and the target which offers-exposes the required capability (and thus the access to the resource).

Figure \ref{fig:capability_routing} shows an example of capability routing in a component instance tree of a generic app. In this example, component \texttt{E} requests access to service \texttt{S} exposed by component \texttt{F}. The request of \texttt{E} is successfully executed since 1) a path between \texttt{F} and \texttt{E} exists in the component instance tree (i.e., arrows 1 to 4), and 2) the path offers-exposes the required service capability for \texttt{S} (\texttt{S} is exposed from \texttt{F} to its parents \texttt{B} and \texttt{A}; \texttt{S} is then offered to \texttt{E} through the child \texttt{C}).

\begin{figure}[!ht]
    \centering
    \includegraphics[width=0.5\textwidth]{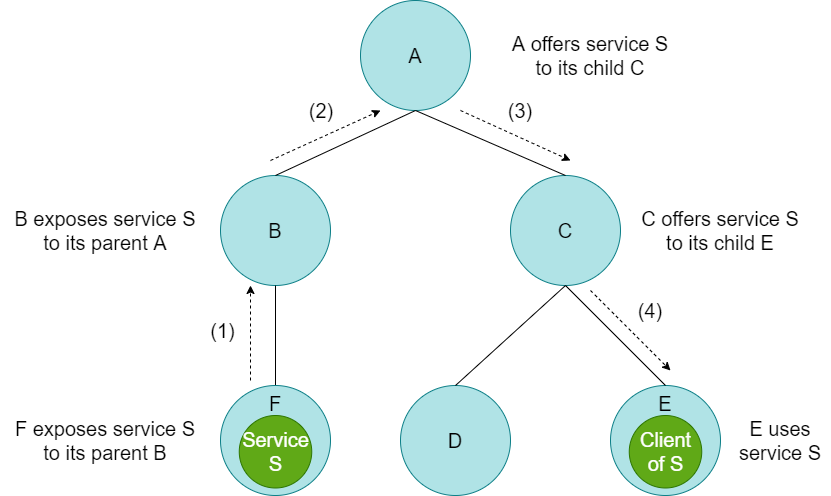}
    \caption{Example of capability routing for the service S.}
    \label{fig:capability_routing}
\end{figure}

\subsection{Package Integrity}

Fuchsia enforces an integrity verification mechanism on software packages. 
In detail, each Fuchsia package is signed using the Merkle Tree algorithm \cite{merkleTree}. The signature information is stored directly inside the package in the meta-information file (called \texttt{meta.far}), similarly to those used for Android APK files \cite{apkSign}. The metadata information includes the package name, the version, and the signature of the content.

To verify the integrity of the package, the Component Framework recomputes the Merkle tree based on the content of the package. If the computed tree matches the one stored in the contents file, i.e., within the \texttt{meta.far}, the package is correctly verified and can be used by the Component Manager to build a new component instance. 

The package signature verification enables Fuchsia to detect undesired modifications in the package archive after it has been signed and to detect corrupts file, e.g., during the download of the package from an external repository.

\section{Discussion and Conclusion} 
\label{sec:discussion}

This paper describes the architecture of Fuchsia OS and its software model, and focuses on the security mechanisms implemented to guarantee a secure execution of apps inside the system. 

Despite being in an early stage of development, Fuchsia exploits a complex set of security features that include sandboxing of processes and components, the use of capabilities, and a specific focus on ICC and on the interaction with the underlying Zircon micro-kernel.
Such design enables Fuchsia to run all the software, including apps and system components, with the least privilege it needs to perform its job and gains access only to the information it needs to know. 

Still, the preliminary security analysis carried out in this work allowed us to unveil some promising research directions that could enhance the overall security of the OS or contribute to the finding of potential security weaknesses.

For instance, the complexity of the layered architecture and the ICC communication mechanisms could lead to unexpected communication flows among the different layers of the Fuchsia stack. 
In particular, we noticed that the Realm Framework allows for the creation of groups of components, called \texttt{collections}, that are hierarchically organized and, thus, can benefit from the capability routing mechanism. In such a case, each capability routed to a component of a collection is automatically extended to all the other members \cite{realm}. This choice could break the principle of least privilege, and lead to grant unnecessary capabilities to one or more components. 

Also, we identified several bottlenecks and critical points in the architecture whose compromise could severely affect the integrity and the availability of the system. A notable example is the Component Manager, which is in charge of handling the lifecycle of components and the capability routing of the entire system.


Finally, we noticed that the package verification mechanisms enable the protection against modifications that would invalidate the signature of the content. As it happens in the Android ecosystem, however, packages are not resilient to repackaging attacks that also involve the re-signing of the tampered file. 
In such a scenario, Fuchsia can be targeted with malicious repackaged apps, as well. 

Thus, future extensions of this work include i) an in-depth analysis of the ICC, with a particular focus on intra-layer and inter-layer communications, and ii) the design of an integrity protection mechanisms for Fuchsia apps that could mitigate repackaging attacks.

\bstctlcite{IEEEexample:BSTcontrol}
\bibliographystyle{IEEEtran}
\bibliography{bibliography}

%
\begin{IEEEbiography}[{\includegraphics[width=1in,height=1.25in,clip,keepaspectratio]{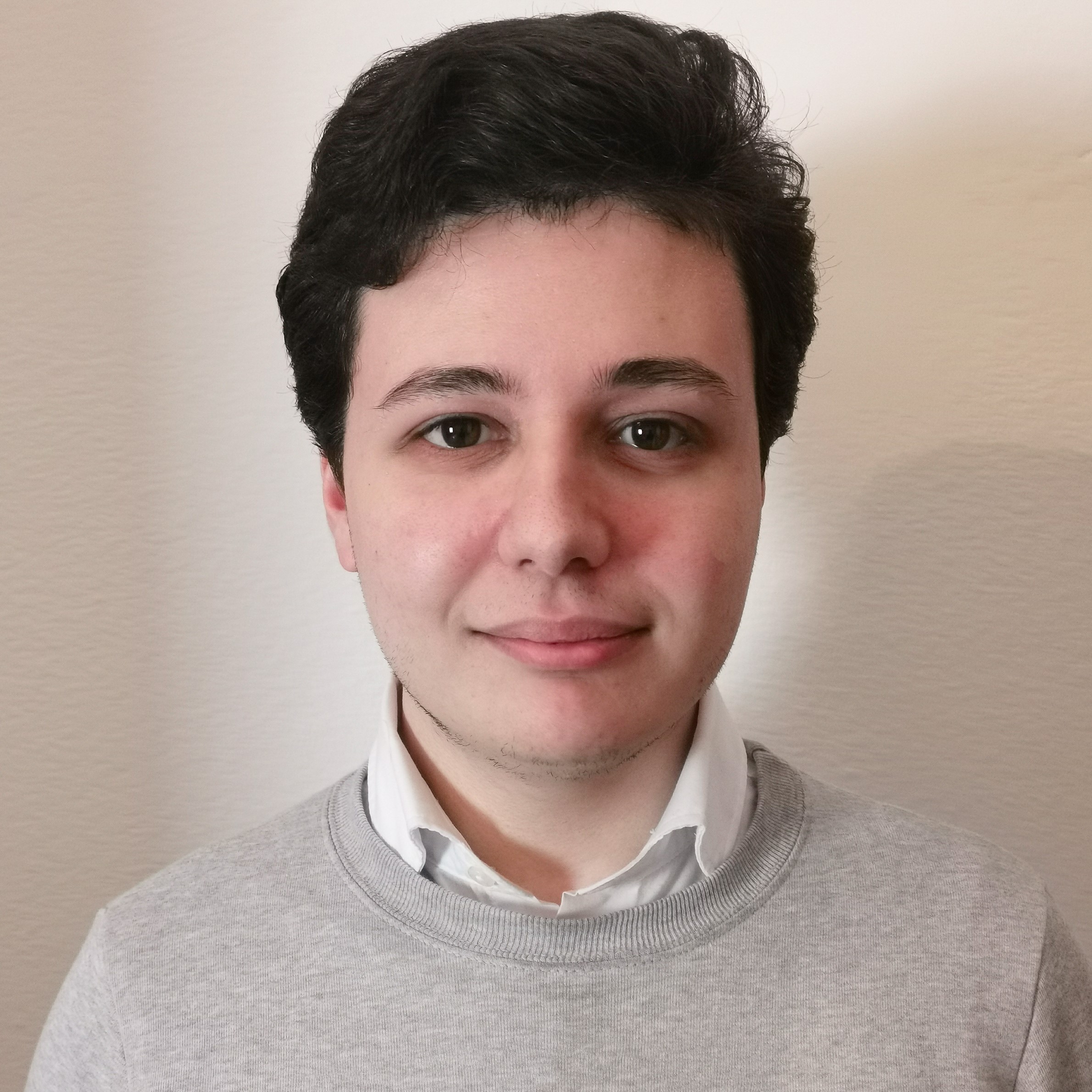}}]{Francesco Pagano} obtained both his BSc and MSc in Computer Engineering at the University of Genoa. His master's degree thesis covered issues on Mobile Privacy, more specifically on the study of information leakages through analytics services, that led to the implementation of the HideDroid app. His thesis, in collaboration with Giovanni Bottino, was supervised by Davide Caputo and Alessio Merlo. Currently, he is working as a post-graduate research fellow at the Computer Security Laboratory (CSEC Lab).
\end{IEEEbiography}


\begin{IEEEbiography}[{\includegraphics[width=1in,height=1.25in,clip,keepaspectratio]{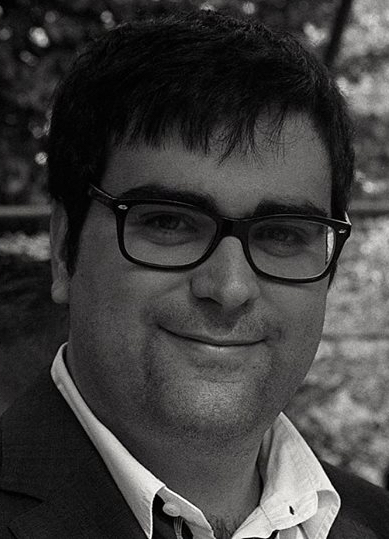}}]{Luca Verderame}obtained his Ph.D. in Electronic, Information, Robotics, and Telecommunication Engineering at the University of Genoa (Italy) in 2016, where he worked on mobile security.
He is currently working as a post-doc research fellow at the Computer Security Laboratory (CSEC Lab), and he is also the CEO and Co-founder of Talos, a cybersecurity startup and university spin-off. His research interests mainly cover information security applied, in particular, to mobile and IoT environments.
\end{IEEEbiography}

\begin{IEEEbiography}[{\includegraphics[width=1in,height=1.25in,clip,keepaspectratio]{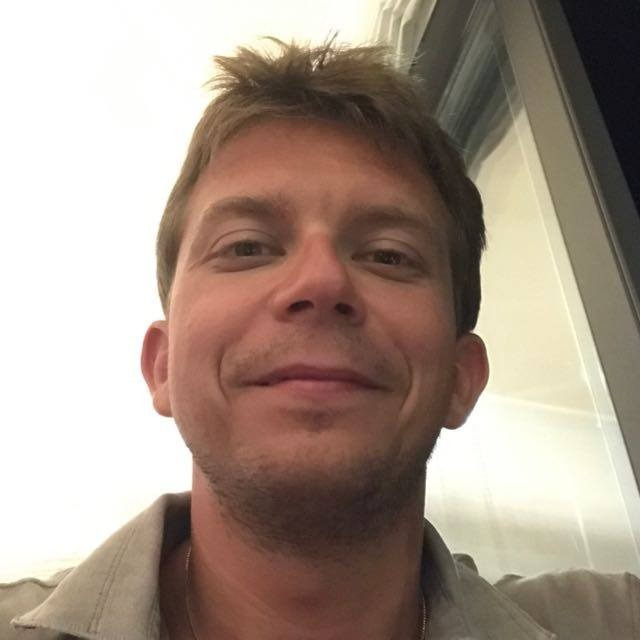}}]{Alessio Merlo} is an Associate Professor in Computer Engineering in the Department of Informatics, Bioengineering, Roboticsand System Engineering Department (DIBRIS) at the University of Genoa, and a member of the Computer Security Laboratory (CSEC Lab). His main research field is Mobile Security, with a specific interest on Android security, automated static and dynamic analysis of Android apps, mobile authentication and mobile malware. More information can be found at: \href{http://csec.it/people/alessio_merlo/}{http://csec.it/people/alessio\_merlo/}.

\end{IEEEbiography}




\end{document}

Email: francesco.pagano@dibris.unige.it